\documentclass[journal]{IEEEtran}

\usepackage{verbatim}  
\usepackage{vector}  
\usepackage{epsf}
\usepackage{epsfig}
\usepackage{epstopdf}
\usepackage{enumerate}
\usepackage{times}
\usepackage{amsmath}
\usepackage{amssymb}
\usepackage{cite}

\usepackage{algorithm}
\usepackage{algpseudocode}

\IEEEoverridecommandlockouts

\usepackage[hidelinks,bookmarks=false]{hyperref}

\title{Low-Complexity Robust MISO Downlink Precoder Design With Per-Antenna Power Constraints}

\author{\IEEEauthorblockN{Mostafa Medra  \quad Timothy N. Davidson\\}

\thanks{This work was supported in part by the Natural Sciences and Engineering Research Council of Canada under grant RGPIN-2015-06631.}
\thanks{A preliminary version of a portion of this work appears in
\textit{Proc. 2015 IEEE Int. Wkshp Signal Process. Adv. Wireless Commun.}}
\thanks{The authors are with the Department of Electrical and Computer Engineering, McMaster University, Hamilton, Ontario, Canada. (E-mail: medramm@mcmaster.ca; davidson@mcmaster.ca).}

}

\begin{document}
\maketitle

\begin{abstract}

This paper considers the design of the beamformers for a multiple-input single-output (MISO) downlink system that seeks to mitigate the impact of the imperfections in the channel state information (CSI) that is available at the base station (BS). The goal of the design is to minimize the outage probability of specified signal-to-interference-and-noise ratio (SINR) targets, while satisfying per-antenna power constraints (PAPCs), and to do so at a low computational cost. Based on insights from the offset maximization technique for robust beamforming, and observations regarding the structure of the optimality conditions, low-complexity iterative algorithms that involve the evaluation of closed-form expressions are developed. To further reduce the computational cost, algorithms are developed for per-antenna power-constrained variants of the zero-forcing (ZF) and maximum ratio transmission (MRT) beamforming directions. In the MRT case, our low-complexity version for systems with a large number of antennas may be of independent interest. The proposed algorithms are extended to systems with both PAPCs and a total power constraint. Simulation results show that the proposed robust designs can provide substantial gains in the outage probability while satisfying the PAPCs.

\end{abstract}


\section{Introduction}

The spatial multiplexing capabilities of base stations (BSs) with multiple antennas offer the potential for substantial gains in the quality of service (QoS) that can be offered to users in a downlink system; e.g., \cite{ShiftingtheMIMO}. In particular, linear beamforming schemes have been developed to simultaneously serve multiple users at their requested signal-to-interference-and-noise ratio (SINR) targets \cite{Reference2,Jointoptimal,Linearprecoding,Iterativemultiuseruplink,OptimalMultiuserTransmit}. However, the performance of those beamforming schemes can be quite sensitive to the accuracy of the channel state information (CSI) that is available at the BS. Since that information is typically obtained through estimation on the uplink (in time division duplexing, TDD, systems) or through estimation on the downlink and quantized feedback (in frequency division duplexing, FDD, systems), the CSI at the BS is inherently uncertain. That observation has spawned the development of a variety of design strategies that incorporate different models for the uncertainty into the design.
One strategy is to require the requested SINR to be met for all channels that are within a specified distance of the BS's model for the channel \cite{Arobustmaximin,ConvexConic,RobustLinear,RobustQoS,Probabilisticallyconstrained,OutageConstrained}.  However, in many scenarios that is a rather conservative approximation of the outage that occurs in practice. Furthermore, although this strategy, or a mild approximation thereof, often results in a convex optimization problem for finding the beamformers, the computational cost of solving those problems can be quite significant. Fortunately, different approaches to approximating the outage probability can yield alternative design strategies that provide excellent performance in practice, even when the uncertainties in the CSI are quite substantial, and do so in a computationally inexpensive way. One such strategy is the offset maximization algorithm \cite{LowComplexityRobustMISO}, in which the beamformers are designed to maximize a carefully structured offset on the performance specification (see Section~\ref{r_algo_subsec}).

The above-mentioned design strategies seek to jointly design the beamforming directions and the power allocated to each direction. However, significant reductions in the computational cost can be obtained by computing the beamforming directions using a (computationally cheap) conventional technique and then developing a robust power loading algorithm. The beamforming directions in this approach are typically chosen to be either the maximum ratio transmission (MRT)~\cite{Maximumratiotransmission} or zero-forcing (ZF) directions \cite{Zeroforcingmethods}. For the case of additive Gaussian uncertainties in the BS's CSI, single-integral expressions for the outage probability can be obtained \cite{OntheDistributionof} and an effective algorithm for finding the power loading that minimizes the power required to meet the specified outage constraint has been developed \cite{Coordinateupdate}. However, that algorithm is rather computationally expensive. In \cite{Peruseroutageconstrained}, insights from bounds on the cumulative distributive function were used to develop a new robust power loading technique that provides performance close to that of the optimal algorithm in \cite{Coordinateupdate}, but has  significantly lower computational cost.

The existing literature on robust downlink beamforming has tended to focus on designs that impose a constraint on the total power transmitted by the BS. In practice, each antenna will typically be driven by its own power amplifier, and hence the design ought to include constraints on the power transmitted from each antenna, as well as the total power. In the case of perfect CSI, a number of downlink beamforming algorithms that incorporate per-antenna power constraints (PAPCs) have been developed \cite{Downlinkbeamformingandpower,OntheDualityoftheMax,Jointdownlink,DownlinkBeamformingUnder,TransmitterOptimization}. For robust beamforming designs that can be formulated as convex problems (e.g.,  \cite{RobustLinear,ConvexConic,OutageConstrained}) and are solved using generic solvers, incorporating these additional constraints is quite straightforward. However, doing so increases the computational cost of what are, in comparison to the perfect CSI case, already quite expensive algorithms. The goal of this paper is to develop robust beamforming designs that incorporate PAPCs and have reasonable computational costs. Our technique is based on insights developed from the offset maximization approach to robust beamforming \cite{LowComplexityRobustMISO}, a closely related power loading technique \cite{Peruseroutageconstrained}, and observations regarding the structure of the optimality conditions for the design problem. These observations enable us to develop a low-complexity dual update optimization startegy related to that in \cite{TransmitterOptimization} that involves the evaluation of a sequence of closed-form expressions. After extending that algorithm to systems that have both PAPCs and a total power constraint, we make the observation that a large fraction of the computational cost arises from the design of the beamforming directions. To reduce that cost, we develop PAPCed variants of the ZF and MRT directions, and show how these can be incorporated into our design approach. Furthermore, we develop a low-complexity version of our PAPCed MRT beamforming algorithm for ``massive MIMO" systems  with a large number of antennas. As scaling techniques for large MRT beamformers have been recently proposed \cite{ModifiedMRT}, that algorithm may be of independent interest.


\section{System model and design approach}

We consider a narrowband  multiple-input single-output (MISO) downlink in which  an $N_t$-antenna BS sends independent messages to $K$ single-antenna users. The transmitted signal at a given signalling instant is constructed using linear beamforming as $\mathbf{x}= \sum_{k=1}^K\mathbf{w}_k s_k$, where $s_k$ is the power-normalized data symbol for user $k$, and $\mathbf{w}_k$ is the associated beamformer. In some settings we will refer to $\mathbf{u}_k=\mathbf{w}_k/ \| \mathbf{w}_k \|$  as the direction of the beamformer, and $\beta_k=\mathbf{w}_k^H\mathbf{w}_k$ as the power allocated to that direction. That enables us to write $$\mathbf{w}_k=\sqrt{\beta_k} \;\mathbf{u}_k.$$ The received signal  at user $k$ can be written as
\begin{equation}\label{rcvd_sig}
    y_k= \mathbf{h}_k^H \mathbf{w}_k s_k + \sum_{j \neq k}\mathbf{h}_k^H \mathbf{w}_j s_j + n_k,
\end{equation}
where $\mathbf{h}_k^H$ denotes the channel between the BS and receiver $k$, and $n_k$ represents the additive zero-mean circular complex Gaussian noise at that user.

In the problems that we will consider, each user specifies the SINR that it will require in order to support the service that it desires. This constraint takes the form
\begin{equation}\label{sinr_constr}
    \text{SINR}_k = \frac{\mathbf{h}_k^H \mathbf{w}_k \mathbf{w}_k^H \mathbf{h}_k}{\mathbf{h}_k^H  (\sum_{j \neq k}\mathbf{w}_j \mathbf{w}_j^H) \mathbf{h}_k + \sigma_k^2} \geq \gamma_k, \\
\end{equation}
where $\sigma_k^2$ is the noise variance at receiver $k$, and $\gamma_k$ is the required SINR. We will find it convenient to rewrite that constraint as $$\mathbf{h}_{k}^H \mathbf{Q}_{k} \mathbf{h}_{k} - \sigma_k^2 \geq 0,$$ where
\begin{equation}\label{Q_eqn}
\mathbf{Q}_{k}=\mathbf{w}_k \mathbf{w}_k^H/\gamma_k-\sum_{j \neq k}\mathbf{w}_j \mathbf{w}_j^H.
\end{equation}

If we denote the signal transmitted from antenna $i$ by $x_i$, then the power constraint on the BS as a whole can be written as $\textstyle\sum_{i=1}^{N_t} E \{|x_i|^2\} =\textstyle\sum_{k=1}^K  \mathbf{w}_k^H \mathbf{w}_k \leq P_t $, where we have used the assumptions that the messages are independent and that the symbols $s_k$ are normalized. If we let $p_i$ denote the  maximum power that can be transmitted from antenna $i$, the PAPC can be written as $E \{|x_i|^2 \} = \bigl[\textstyle\sum_{k=1}^K \mathbf{w}_k \mathbf{w}_k^H \bigr]_{i,i}  \leq p_i$, where $[\cdot]_{i,i}$ denotes the ($i$,$i$)th entry of the given matrix.

In order for a BS to be able to evaluate whether a candidate set of beamformers $\{\mathbf{w}_k\}_{k=1}^K$ satisfies the $K$ SINR constraints in \eqref{sinr_constr}, the  BS must know each channel vector $\mathbf{h}_k$; e.g.,\cite{Reference2}. However, typically the BS will only have an estimate of each channel, denoted $\mathbf{h}_{e_k}$. To incorporate the uncertainty in that channel estimate into the design, we will postulate a conditional distribution, $p(\mathbf{h}_k|\mathbf{h}_{e_k})$, and convert the deterministic QoS constraint $ \text{SINR}_k \geq \gamma_k$ into the chance constraint $\text{Prob}(\text{SINR}_k \geq \gamma_k)\geq 1- \delta_k$, where  $\delta_k$ is the required outage probability. In this paper, we will model the uncertainty additively; i.e.,
\begin{equation}\label{uncertainty}
   \mathbf{h}_{k}=\mathbf{h}_{e_{k}}+\mathbf{e}_{k},
\end{equation}
with $\mathbf{e}_{k}$ having zero-mean and being independent of the channel and data. Our results will focus on the case where $\mathbf{e}_{k}$ is a zero-mean circular Gaussian random variable of covariance $\sigma_{e_{k}}^2 \mathbf{I}$. Among a number of scenarios, that model is appropriate in certain TDD systems in which channels are estimated during the uplink training phase.

\subsection{Design approach}
With the uncertainty modeled as described above, one approach to the design of the downlink beamformers $\mathbf{w}_k$ is to seek to minimize the probability of outage of the SINR targets, subject to  a total power constraint and PAPCs; i.e.,
\begin{subequations}\label{outage_min}
\begin{align}
    \min_{\substack{\mathbf{w}_k, \delta_k}} \    \max_{\substack{k}} \quad & \delta_k \\
   \text{s.t.} \quad   & \textstyle\sum_{k}  \mathbf{w}_k^H \mathbf{w}_k  \leq P_t, \\
    \ & \Bigl[\textstyle\sum_{k=1}^K \mathbf{w}_k \mathbf{w}_k^H \Bigr]_{i,i}\leq   p_i, \quad \forall i,\\
   \quad  & \text{Prob}(\text{SINR}_k \geq \gamma_k)\geq 1- \delta_k, \quad \forall k. \label{sinr2}
     \end{align}
\end{subequations}
This problem is hard to solve even without the PAPCs. However, in the case that the PAPCs are omitted, the offset maximization algorithm \cite{LowComplexityRobustMISO} is a low-complexity algorithm that has been shown to provide good performance. The goal of this paper is to use insights from the development of the offset maximization approach  to develop an effective low-complexity algorithm for the PAPCed case.  One observation that we will use is that the performance of the offset maximization approach  can be improved by applying the robust power loading algorithm in \cite{Peruseroutageconstrained} to the beamforming directions generated by the offset maximization.
Doing so reveals that robust beamformers can be obtained with a computational cost that is similar to that of beamformer design in the perfect CSI case. (Many existing approaches to robust beamforming are much more expensive than the perfect CSI case; e.g., \cite{Probabilisticallyconstrained,OutageConstrained}.) However, like the perfect CSI case, it is the computation of the directions that dominate the computational cost. Therefore, we also propose to apply the principles that underlie the power loading in \cite{Peruseroutageconstrained} to beamforming directions that can be computed more efficiently, such as PAPCed variants, derived herein, of the classical ZF and MRT directions; see Sections~\ref{sect_ZF}, and \ref{sect_MRT}. In the latter case, a further approximation that is suitable for scenarios with a large number of antennas at the BS substantially reduces the computational cost, and has almost the same outage performance.

To lay the groundwork for the development of the proposed beamforming schemes, in the following subsections we briefly review the offset maximization approach to beamformer design under a total power constraint \cite{LowComplexityRobustMISO}, and the low-complexity robust power loading technique for systems with a total power constraint that was developed in \cite{Peruseroutageconstrained}.

%

\subsection{Offset maximization beamforming directions} \label{r_algo_subsec}

The offset maximization beamformers \cite{LowComplexityRobustMISO} can be found by solving the following problem:
\begin{subequations}\label{r_prob}
\begin{align}
     r_{t}^\star=\max_{\substack{\mathbf{w}_k}, r} \quad & r \\
    \text{s.t.} \quad &\textstyle\sum_{k=1}^K  \mathbf{w}_k^H \mathbf{w}_k \leq P_t, \label{r_prob_p} \\
    &\mathbf{h}_{e_k}^H \mathbf{Q}_{k} \mathbf{h}_{e_k} - \sigma_k^2-r \geq 0, \quad \forall k. \label{sinr3}
    \end{align}
\end{subequations}
It is implicit in \eqref{sinr3} that this algorithm tries to find the largest noise-plus-interference power each user can endure, under the total power constraint. In \cite{LowComplexityRobustMISO} an efficient method to solve \eqref{r_prob} was developed by considering the following problem, in which, for now, it is assumed that the optimal value for \eqref{r_prob}, $r_{t}^\star$, is known:
\begin{subequations}\label{r_prob2}
\begin{align}
      P^\star=\min_{\substack{\mathbf{w}_k}} \quad  &\textstyle\sum_{k}  \mathbf{w}_k^H \mathbf{w}_k \\
       \text{s.t.} \quad &\mathbf{h}_{e_k}^H \mathbf{Q}_{k} \mathbf{h}_{e_k} - \sigma_k^2-r_{t}^\star \geq 0, \quad \forall k. \label{sinr1}
      \end{align}
\end{subequations}
It can be shown \cite{LowComplexityRobustMISO} that the optimal value of the problem in \eqref{r_prob2} is $P_t$, and that  any set of beamformers that optimize  \eqref{r_prob2} are also optimal for \eqref{r_prob}. Also, at optimality, all the constraints are satisfied with equality.

The advantage of the connection between problems \eqref{r_prob} and \eqref{r_prob2} is that a highly efficient algorithm for the problem in \eqref{r_prob2} with  $r=0$ (i.e., the perfect CSI case) was developed in \cite{Jointoptimal}; see also \cite{OptimalMultiuserTransmit}. That algorithm can be extended to jointly find the optimal beamformers and the optimal  offset,  $r_{t}^\star$, for the problem in \eqref{r_prob}.  In particular, if we let $\nu_k$  denote the Lagrange multiplier for the SINR constraint in \eqref{sinr1}, then from the KKT conditions of \eqref{r_prob2} we can find the offset maximization directions by solving the  eigen problem
\begin{equation}\label{closed_form}
\mathbf{u}_k =\Biggl( \frac{\nu_k}{\gamma_k}\mathbf{h}_{e_k} \mathbf{h}_{e_k}^H-\sum_{j\neq k} \nu_j \mathbf{h}_{e_j} \mathbf{h}_{e_j}^H \Biggr)\mathbf{u}_k,
\end{equation}
where the Lagrange multipliers must satisfy the fixed-point relation
\begin{equation}\label{nu}
\nu_k^{-1}  = \mathbf{h}_{e_k}^H \Bigl(\mathbf{I}_{N_t}+\textstyle\sum_{j} \nu_j \mathbf{h}_{e_j} \mathbf{h}_{e_j}^H \Bigr)^{-1} \mathbf{h}_{e_k}  \Bigl(1+\frac{1}{\gamma_k} \Bigr).
\end{equation}
Since \eqref{closed_form} can be solved using a power method, the complexity of finding the directions is dominated by the matrix inversion in \eqref{nu}, which requires $\mathcal{O}(N_t^3)$ operations. Having found those directions, the offset maximization power loading and the optimal offset can be found by solving the $K+1$ linear equations that arise when the constraints in \eqref{r_prob_p} and \eqref{sinr3} hold with equality.

\subsection{Robust power loading} \label{power_loading_algo}

The offset maximization algorithm described above uses the same offset $r$ to increase the robustness of each user to channel uncertainty. The goal of robust power loading approach in \cite{Peruseroutageconstrained} is to provide a computationally-efficient way to adapt the offset to the characteristics of each user's channel. For an arbitrary set of beamforming directions $\{ \mathbf{u}_k\}$, the generic power loading problem can be stated as
\begin{subequations}\label{outage_min}
\begin{align}
    \min_{\substack{\beta_k, \delta_k}}   \quad &  \max_{\substack{k}}  \quad  \delta_k \\
    \text{s.t.}  \quad   &\sum_{k=1}^K \beta_k \leq P_t, \\
     & \text{Prob}(\text{SINR}_k \geq \gamma_k)\geq 1- \delta_k, \quad \forall k.  \label{sinr}
     \end{align}
\end{subequations}

The derivation of the algorithm developed in \cite{Peruseroutageconstrained} for producing good solutions to \eqref{outage_min} begins by observing the under the additive uncertainty model in \eqref{uncertainty}, the probability  that $\text{SINR}_k \geq \gamma_k$ is equal to the probability that
\begin{equation}\label{SINR_reformulation}
   f_k(\mathbf{e}_{k})=\mathbf{h}_{e_k}^H \mathbf{Q}_{k} \mathbf{h}_{e_k} + 2 \text{Re}(\mathbf{e}_{k}^H  \mathbf{Q}_{k} \mathbf{h}_{e_k}  ) + \mathbf{e}_{k}^H  \mathbf{Q}_{k} \mathbf{e}_{k}  - \sigma_k^2 \geq 0.
\end{equation}
If we assume that the norms of the errors $\mathbf{e}_{k}$ are small, as they will need to be for reliable operation \cite{MIMObroadcast}, then  we can approximate the quadratic term $ \mathbf{e}_{k}^H  \mathbf{Q}_{k} \mathbf{e}_{k}$ by a Gaussian random variable of the same mean and  variance. In that case, the distribution of $f_k(\mathbf{e}_{k})$ becomes Gaussian. (Recall that we are focusing on the case where $ \mathbf{e}_{k}$ is Gaussian, with zero mean and of covariance $\sigma_{e_{k}}^2 \mathbf{I}$; cf. \eqref{uncertainty}.) Under that approximation,  if we design the power loading so that the mean, $\mu_{f_k}$, of $f_k(\mathbf{e}_{k})$  is a significant multiple of its standard deviation, $\sigma_{f_k}$, then that user will achieve a low outage probability.
Indeed, we can choose a value for that multiple so that the target outage probability is guaranteed to be satisfied; see, e.g., \cite{Arobustmaximin}. We also note that the optimal solution of \eqref{outage_min} has equal values for $\delta_k$. If that were not the case, the user(s) with higher outage probability could be allocated more power and the other user(s) less, which would reduce the objective value, and thus contradict the assumed optimality. Therefore, it is natural to choose the same multiple, $r$, for each user in the approximation of the outage constraint in \eqref{sinr}.
The resulting approximation of the problem in \eqref{outage_min} can be written as \cite{Peruseroutageconstrained}
\begin{subequations}\label{outage_min2}
\begin{align}
    \max_{\substack{\beta_k, r}}   \quad & r \\
     \text{s.t.}  \quad   &\sum_{k=1}^K \beta_k \leq P_t, \label{outage_min2_pc} \\
    & \mu_{f_k} \geq r \sigma_{f_k}, \quad \forall k. \label{sinr2}
    \end{align}
\end{subequations}
From the definition of $f_k(\mathbf{e}_{k})$ in \eqref{SINR_reformulation} and the channel uncertainty model in \eqref{uncertainty}, it can be shown that
\begin{equation}\label{mean_eqn}
\begin{aligned}
 \mu_{f_k}= \mathbf{h}_{e_k}^H \mathbf{Q}_{k} \mathbf{h}_{e_k}- \sigma_k^2  + \sigma_e^2 \beta_k \left( 1/\gamma_k +1 \right)- \sigma_e^2 P_t,
   \end{aligned}
\end{equation}
which is linear in the design variables $\{\beta_k \}_{k=1}^K$. (Recall from  \eqref{Q_eqn} that $\mathbf{Q}_{k}=\beta_k \mathbf{u}_k \mathbf{u}_k^H/\gamma_k-\sum_{j \neq k} \beta_j \mathbf{u}_j \mathbf{u}_j^H.$) Similarly, we have that
\begin{equation}\label{var_rel}
\begin{aligned}
     \sigma_{f_k}^2=\text{var}\{f_k(\mathbf{e}_{k}) \} =2 \sigma_e^2 \mathbf{h}_{e_k}^H  \mathbf{Q}_{k}^2 \mathbf{h}_{e_k} +\sigma_e^4 \text{tr} (\mathbf{Q}_{k}^2).
   \end{aligned}
\end{equation}

The structure of the problem in \eqref{outage_min2} is such that the constraints hold with equality at optimality \cite{Peruseroutageconstrained}. Since $ \sigma_{f_k}$ is not a linear function  in $\boldsymbol{\beta}$, that results in a set of non-linear equations for the power loading. The following iterative linearization technique has been shown in \cite{Peruseroutageconstrained} to be an effective way to obtain good solutions to \eqref{outage_min2}:
\begin{enumerate}
  \item Initialize each $\sigma_{f_k}=1$.
  \item Find $\{\beta_k\}$ and $r$ by solving the set of linear equations that arise from equality in  \eqref{outage_min2_pc} and \eqref{sinr2} for the current values of $\sigma_{f_k}$, where $\mu_{f_k}$ is defined in \eqref{mean_eqn}.
  \item Update each $\sigma_{f_k}$ using \eqref{var_rel}.
  \item Return to (2) until a  convergence criterion is satisfied.
\end{enumerate}
We note that the matrix that relates $\{\beta_k\}$ to $\sigma_{f_k}$ and $r$ in step 2 is constant, and, accordingly, we need only invert this matrix once \cite{Peruseroutageconstrained}. In practice, this algorithm converges quickly with a high probability \cite{Peruseroutageconstrained}.
In \cite{RobustMISOdownlinkAnefficient}, the performance of this algorithm was shown to provide very similar performance to the optimal power loading in \cite{Coordinateupdate}, and at a cost that is dominated by the $\mathcal{O}(K^3)$ operations that result from the initial matrix inversion.

\section{Offset maximization designs with PAPCs }
To simplify the development of the proposed robust beamforming technique, we will first consider the addition of PAPCs to the offset maximization problem in \eqref{r_prob}. We will then modify the resulting algorithm using insights from the above robust power loading algorithm.

When we add the PAPCs to the offset maximization problem in \eqref{r_prob}, the design problem becomes
\begin{subequations}\label{pr_gen}
\begin{align}
     r_{tpa}^\star=\max_{\substack{\mathbf{w}_k}, r} \quad & r \\
    \text{s.t.} \quad&\textstyle\sum_{k=1}^K  \mathbf{w}_k^H \mathbf{w}_k \leq P_t,\\
    \ & \Bigl[\textstyle\sum_{k=1}^K \mathbf{w}_k \mathbf{w}_k^H \Bigr]_{i,i}\leq   p_i, \quad \forall i,\\
    &\mathbf{h}_{e_k}^H \mathbf{Q}_{k} \mathbf{h}_{e_k} - \sigma_k^2-r \geq 0, \quad \forall k. \label{pr_gen_sinr}
    \end{align}
\end{subequations}
Although the formulation in \eqref{pr_gen} is not convex, it can be transformed in a straightforward way into a second order cone program, using the technique that was used for the case of  perfect CSI; cf. \cite{Linearprecoding,TransmitterOptimization}. While that formulation can be solved using a generic interior point method (e.g., \cite{Reference1}), such generic methods do not exploit the structure of the problem, and the development of tailored algorithms that do exploit the structure  offers the potential for improved computational efficiency.

In the following subsections, we will first develop a low-complexity algorithm for the case where we have  PAPCs only, with no total power constraint. Then we will tackle the general problem with both types of power constraints. The development will use insights from algorithms developed for the perfect CSI case \cite{TransmitterOptimization} and insights from the robust power loading algorithm described in Section~\ref{power_loading_algo}.

\subsection{Dominant PAPCs}

If $P_t > \sum_{i=1}^{N_t} p_i$, the total power constraint can never be active and the problem in \eqref{pr_gen} can be rewritten as
\begin{subequations}\label{pr_gen2}
\begin{align}
     r_{pa}^\star=\max_{\substack{\mathbf{w}_k}, r} \quad & r \\
    \text{s.t.} \quad & \Bigl[\textstyle\sum_{k=1}^K \mathbf{w}_k \mathbf{w}_k^H \Bigr]_{i,i}\leq p_i, \quad \forall i,\label{pr_gen2_2} \\
    &\mathbf{h}_{e_k}^H \mathbf{Q}_{k} \mathbf{h}_{e_k} - \sigma_k^2-r \geq 0, \quad \forall k.\label{pr_gen2_3}
    \end{align}
\end{subequations}
Motivated by the way that a customized algorithm for \eqref{r_prob2} was adapted \cite{LowComplexityRobustMISO} to solve the problem in  \eqref{r_prob}, we consider the following problem in which, for now,  $r_{pa}^\star$ is presumed to be known,
\begin{subequations}\label{pr_gen3}
\begin{align}
      \min_{\substack{\mathbf{w}_k}, \alpha} \quad & \alpha \textstyle\sum_{i=1}^{N_t} p_i\\
   \text{s.t.} \quad & \Bigl[\textstyle\sum_{k=1}^K \mathbf{w}_k \mathbf{w}_k^H \Bigr]_{i,i}\leq \alpha p_i, \quad \forall i,\label{pr_gen3_2} \\
    &\mathbf{h}_{e_k}^H \mathbf{Q}_{k} \mathbf{h}_{e_k} - \sigma_k^2-r_{pa}^\star \geq 0, \quad \forall k. \label{pr_gen3_3}
    \end{align}
\end{subequations}
In the context of \eqref{pr_gen3}, the constant term $\sum_{i=1}^{N_t} p_i$ in the objective is superfluous, but it will simplify the interpretation of the Lagrangian. Using arguments analogous to those in \cite{LowComplexityRobustMISO,TransmitterOptimization},  it can be shown that any set of beamformers that is optimal for  \eqref{pr_gen3} is also optimal for \eqref{pr_gen2}, and the optimal value of $\alpha$ in \eqref{pr_gen3} is one.

Now, let $q_i$ denote the dual variable of the $i$th condition in \eqref{pr_gen3_2} and $\nu_k$ denote the dual variable of the $k$th condition in \eqref{pr_gen3_3}. Let us also define the diagonal matrix $\hat{\mathbf{Q}}$, such that $[\hat{\mathbf{Q}}]_{i,i}=q_i$. These definitions enable us to write the Lagrangian of the problem in \eqref{pr_gen3} as
\begin{multline}
     \mathcal{L}(\mathbf{w}_k,\alpha,\nu_k,q_i)=\sum_{k=1}^{K}\nu_k (\sigma_k^2+r_{pa}^\star) + \alpha \Bigl( \sum_{i=1}^{N_t} p_i- \sum_{i=1}^{N_t}q_i  p_i \Bigr) \\ +\sum_{k=1}^{K}\mathbf{w}_k^H \Bigl(\hat{\mathbf{Q}}+\sum_{j \neq k}\nu_j \mathbf{h}_{e_j} \mathbf{h}_{e_j}^H -\nu_k/\gamma_k \mathbf{h}_{e_k} \mathbf{h}_{e_k}^H  \Bigr) \mathbf{w}_k.
\end{multline}
Using the notion of complementary slackness, since the optimal value of $\alpha$ is one, at optimality we have that  $\sum_{i=1}^{N_t} p_i- \sum_{i=1}^{N_t}q_i  p_i = 0$. Also, at optimality we have $\hat{\mathbf{Q}}+\sum_{j \neq k}\nu_j \mathbf{h}_{e_j} \mathbf{h}_{e_j}^H -\nu_k/\gamma_k \mathbf{h}_{e_k} \mathbf{h}_{e_k}^H \succeq \mathbf{0}$, with $\mathbf{w}_k$ lying in the null space of this matrix. This can be simplified to show that  $\mathbf{w}_k$ and
\begin{equation}\label{weqn}
\hat{\mathbf{w}}_k =\Bigl(\hat{\mathbf{Q}}+\sum_{k}\nu_k \mathbf{h}_{e_k} \mathbf{h}_{e_k}^H \Bigr)^{\dag}  \mathbf{h}_{e_k},
\end{equation}
where $(\cdot)^\dag$ denotes the Moore-Penrose pseudo-inverse,
should be in the same direction. Further simplifications show that the dual variable $\nu_k$ in \eqref{weqn} should satisfy the fixed point equation
\begin{equation}\label{lambda}
\nu_k^{-1}  = \mathbf{h}_{e_k}^H \Bigl(\hat{\mathbf{Q}}+\textstyle\sum_{j} \nu_j \mathbf{h}_{e_j} \mathbf{h}_{e_j}^H \Bigr)^{\dag} \mathbf{h}_{e_k}  \Bigl(1+\frac{1}{\gamma_k} \Bigr).
\end{equation}

From \eqref{lambda} we observe that if we were given  the optimal $\hat{\mathbf{Q}}$, we could find the optimal values for $\{\nu_k \}$ using \eqref{lambda} and then the optimal directions $ \{\mathbf{u}_k \}$ by normalizing the $ \{\hat{\mathbf{w}}_k \}$ obtained using \eqref{weqn}. After doing so, we could complete the solution of \eqref{pr_gen3} by finding the optimal values for $\beta_k=\|\mathbf{w}_k\|^2$. That can be done by solving the set of $K$ linear equations that arise from the fact that at optimality \eqref{pr_gen3_3} holds with equality. (If this were not the case for condition $k$ in  \eqref{pr_gen3_3}, then the amplitude of $\mathbf{w}_k$ could be decreased which would allow a smaller value of $\alpha$ while satisfy all the other constraints.) To adapt that approach to solve \eqref{pr_gen2}, in the final step we must simultaneously solve for $\{\beta_k\}$ and $r_{pa}^\star$. To do so we observe that $r_{pa}^\star$ enters linearly into \eqref{pr_gen3_3}, and hence all we need is one more linearly independent equation. To obtain that equation we observe that if $q_i>0$, then the $i$th component of \eqref{pr_gen3_2} holds with equality. By summing over all the active constraints in \eqref{pr_gen3_2} we obtain the following equation
\begin{equation}\label{extra_eqn}
\sum_{i, \forall q_i\neq 0} \Bigl[\textstyle\sum_{k=1}^K  \beta_k \mathbf{u}_k \mathbf{u}_k^H \Bigr]_{i,i}= \textstyle\sum_{i, \forall q_i\neq 0}  p_i.
\end{equation}
In the case that all the $q_i$ are positive --- a case that happens quite often --- the equation in \eqref{extra_eqn} simplifies to $\textstyle\sum_{k=1}^K  \beta_k = \textstyle\sum_{i=1}^{N_t}  p_i.$
%

To complete the algorithm, we need to develop a technique to determine the optimal $\hat{\mathbf{Q}}$. One strategy for doing so is to apply the projected subgradient technique developed in~\cite{TransmitterOptimization}. That involves applying the update equation  $\hat{\mathbf{Q}}^{n+1}=\text{proj} \bigl( \hat{\mathbf{Q}}^{n}+ t_n \text{diag} (\text{diag}(\textstyle\sum_{i} \mathbf{w}_i \mathbf{w}_i^H)) \bigr)$, where proj($\cdot$) denotes the projection of a matrix on the space of diagonal positive semidefinite matrices that satisfy $\sum_{i=1}^{N_t}q_i  p_i=\sum_{i=1}^{N_t} p_i$ and, consistent with the syntax used in \textsc{Matlab}, when diag($\cdot$) operates on a matrix it produces a vector containing the diagonal elements and when it operates on a vector it produces a diagonal matrix with the elements of the vector on the diagonal. The initialization parameters used in~\cite{TransmitterOptimization} were chosen to be $\hat{\mathbf{Q}}^{0}=\mathbf{I}$ and the step size chosen to be $t_n=1/n$. Although this strategy converges, it can be quite slow~\cite{TransmitterOptimization}. In this paper, we will refine the approach in two ways. First, in Appendix~\ref{q_update_app} we develop a computationally cheap quasi-closed-form expression for the projection of $\hat{\mathbf{Q}}^{n+1}$ in a 2-norm sense.
Second, based on insights from \cite{DecompositionbyPartialLinearization} we will choose a step size of the form $t_n=t_{n-1}-t_{n-1}^2/a$, for some positive scalar $a$. In addition, in Section~\ref{algo_acc} we will identify a prediction step that can be used in the first iteration to accelerate the algorithm. One simple termination strategy is to stop the algorithm when $ [\sum_{k} \mathbf{w}_k \mathbf{w}_k^H]_{i,i}-p_i < \epsilon_i, \forall i$, where $\epsilon_i$ is the maximum allowable violation of the  power constraint for the $i$th antenna.
Following the above development, the algorithm can be summarized as shown in Algorithm \ref{Alg1}.

 \begin{algorithm}
\caption{Offset maximization with PAPCs}
\label{Alg1}
\begin{algorithmic}[1]
\State Initialize the diagonal matrix $\hat{\mathbf{Q}}^{0}$ such that each element is non-negative and $\sum_{i=1}^{N_t}q_i  p_i=\sum_{i=1}^{N_t} p_i$. Set $n=0$.
\While {$[\sum_{k} \mathbf{w}_k \mathbf{w}_k^H]_{i,i}-p_i>\epsilon_i$ for any $i$}
        \State Find $\{\nu_k\}$ using \eqref{lambda}.
  \State Solve for the directions $\{\mathbf{u}_k\}$ by normalizing the $\{\hat{\mathbf{w}}_k\}$ obtained using \eqref{weqn}.
  \State Find the power loading $\{ \beta_k\}$ and $r_{pa}^\star$ by solving the set of linear equations arising from \eqref{pr_gen3_3} holding with equality and \eqref{extra_eqn}.
  \State Update $\hat{\mathbf{Q}}^{n+1}$ using the results in Appendix~\ref{q_update_app}.
  \State Increment $n$.
\EndWhile
\end{algorithmic}
\end{algorithm}

Having developed an efficient algorithm for the offset maximization problem with PAPCs, we now seek to incorporate the principles of the robust power loading discussed in  Section~\ref{power_loading_algo}. To do so, we note that in the offset maximization design, the directions are independent of the offset term $r$ in \eqref{pr_gen2_3}; cf.  \eqref{weqn} and  \eqref{lambda}. That suggests that we could simply modify the power loading step. Indeed, once the directions have been obtained in step 4 of Algorithm~\ref{Alg1}, we can replace the power loading in step 5  by the $\{ \beta_k \}$ and  $r^\star$ that solve \eqref{outage_min2}. Those values can be found using the algorithm in  Section~\ref{power_loading_algo}; see \cite{Peruseroutageconstrained}. Incorporating that robust power loading algorithm into the framework of Algorithm~\ref{Alg1} results in  Algorithm~\ref{Alg2}.

\begin{algorithm}
\caption{PAPCed offset maximization with robust power loading}
\label{Alg2}
\begin{algorithmic}[1]
\State Initialize the diagonal matrix $\hat{\mathbf{Q}}^{0}$ such that each element is non-negative and $\sum_{i=1}^{N_t}q_i  p_i=\sum_{i=1}^{N_t} p_i$. Set $n=0$.
\While {$[\sum_{k} \mathbf{w}_k \mathbf{w}_k^H]_{i,i}-p_i>\epsilon_i$ for any $i$}
 \State Find $\{\nu_k\}$  using \eqref{lambda}.
  \State Solve for the directions $\{\mathbf{u}_k\}$ by normalizing  $\{\hat{\mathbf{w}}_k\}$ obtained using \eqref{weqn}.
   \State Find $\{\beta_k\}$ and $r^\star$ by solving $\mathbb{E} (\mathbf{h}_{k}^H \mathbf{Q}_{k} \mathbf{h}_{k} - \sigma_{k}^2 )= \sigma_{s_k} r^\star$ and \eqref{extra_eqn} using the method provided in Section~\ref{power_loading_algo}.
  \State Update $\hat{\mathbf{Q}}^{n+1}$ using the results in Appendix~\ref{q_update_app}.
  \State Increment $n$.
\EndWhile
\end{algorithmic}
\end{algorithm}

\subsection{Total and PAPCed algorithm}\label{gen_offset_max}

Using the principles outlined in Section~\ref{r_algo_subsec} and the previous section, we can develop an algorithm for solving the general problem in \eqref{pr_gen}, which has PAPCs and a total power constraint. In this section, we will focus on the case when $P_t$ is sufficiently smaller than $\sum_i p_i$ to ensure that the total power constraint is active. (Otherwise, the problem can be solved by the techniques in the previous section.)  Similar to the previous section, we will obtain the beamforming directions by normalizing the beamformers resulting from the following problem
\begin{subequations}\label{r_prob_gen_eq}
\begin{align}
     \min_{\substack{\mathbf{w}_k}} \quad & \textstyle\sum_{k=1}^K \mathbf{w}_k^H \mathbf{w}_k \\
    \text{s.t.} \quad  & \Bigl[\textstyle\sum_{j=1}^K \mathbf{w}_j \mathbf{w}_j^H \Bigr]_{i,i}\leq   p_i,  \quad \forall  i\\
   & \mathbf{h}_{e_{k}}^H \mathbf{Q}_{k} \mathbf{h}_{e_{k}} - \sigma_{k}^2-r_{tpa}^\star \geq 0, \quad \forall k,
   \end{align}
\end{subequations}
and then we will refine  the power loading using the method described in Section~\ref{power_loading_algo}.
As in the previous development, the Lagrangian of \eqref{r_prob_gen_eq} plays a key role. It can be written as
\begin{multline}
     \mathcal{L}(\mathbf{w}_k,\nu_k,q_i)=\sum_{k=1}^{K}\nu_k (\sigma_k^2+r_{tpa}^\star) - \sum_{i=1}^{N_t}q_i  p_i + \\ \sum_{k=1}^{K}\mathbf{w}_k^H \Bigl(\mathbf{I}+\hat{\mathbf{Q}}+\sum_{j \neq k}\nu_j \mathbf{h}_{e_j} \mathbf{h}_{e_j}^H -\nu_k/\gamma_k \mathbf{h}_{e_k} \mathbf{h}_{e_k}^H  \Bigr) \mathbf{w}_k.
\end{multline}
Using the KKT conditions, for a given value for $\hat{\mathbf{Q}}$ we can compute the corresponding directions and then the robust power loading in Section~\ref{power_loading_algo}. Furthermore, the subgradient used in the previous section remains a subgradient in this case. However, the structure of the KKT conditions is simpler in this case, which results in a more straightforward projection for the $\hat{\mathbf{Q}}$ matrix. Indeed, since the only constraint on $q_i$ in this case is that it is non-negative, the update equation for $\hat{\mathbf{Q}}$ can be written as
\begin{equation}\label{qupdate_gen}
 \hat{\mathbf{Q}}^{n+1}=\max\Bigl( \hat{\mathbf{Q}}^{n}+ t_n \text{diag} \Bigl(\text{diag}\Bigl(\sum_{k} \mathbf{w}_k \mathbf{w}_k^H \Bigr)- \mathbf{p} \Bigr), \mathbf{0}\Bigr),
\end{equation}
where the maximum operator is defined element-wise, and $\mathbf{p}$ is the vector whose $i$th element is $p_i$.
Therefore, we can construct an algorithm that has a similar structure to that in Algorithm~\ref{Alg2}. Having said that, in the case of PAPCs only there is a strong likelehood that the PAPCs will be active at optimality, and hence it makes sense to initialize the algorithm with a positive definite matrix $\hat{\mathbf{Q}}^0$. In the general case, the PAPCs are less likely to be active at optimality, and hence we will initialize the algorithm with $\hat{\mathbf{Q}}^0=\mathbf{0}.$ The resulting algorithm is provided in Algorithm~\ref{Alg3}.

\begin{algorithm}
\caption{Generalized offset maximization}
\label{Alg3}
\begin{algorithmic}[1]
\State Initialize $\hat{\mathbf{Q}}^{0}=\mathbf{0}$. Set $n=0$.
\While {$[\sum_{k} \mathbf{w}_k \mathbf{w}_k^H]_{i,i}-p_i>\epsilon_i$ for any $i$}
 \State Find $\nu_k$ using the fixed point equations \newline $\nu_k^{-1}  = \mathbf{h}_{e_k}^H \Bigl(\mathbf{I}+\hat{\mathbf{Q}}^n+\textstyle\sum_{j} \nu_j \mathbf{h}_{e_j} \mathbf{h}_{e_j}^H \Bigr)^{-1} \mathbf{h}_{e_k}  \Bigl(1+1/\gamma_k \Bigr).$
  \State Solve for the directions $\mathbf{u}_k=\hat{\mathbf{w}}_k/ \| \hat{\mathbf{w}}_k \|$, where \newline $\hat{\mathbf{w}}_k =\Bigl(\mathbf{I}+\hat{\mathbf{Q}}^n+\sum_{j}\nu_j^n \mathbf{h}_{e_j} \mathbf{h}_{e_j}^H \Bigr)^{-1}  \mathbf{h}_{e_k}$.
  \State Find $\{\beta_k\}$ and $r^\star$  by solving $\mathbb{E} (\mathbf{h}_{k}^H \mathbf{Q}_{k} \mathbf{h}_{k} - \sigma_{k}^2 )= \sigma_{s_k} r^\star$ and $\sum_k \beta_k = P_t$ using the method provided in Section~\ref{power_loading_algo}.
  \State Update $\hat{\mathbf{Q}}^{n+1}$ using \eqref{qupdate_gen}.
  \State Increment $n$.
\EndWhile
\end{algorithmic}
\end{algorithm}

\subsection{Algorithm acceleration}\label{algo_acc}

As will be apparent in the simulations in Section~\ref{sect_sim}, the modified update in Appendix~\ref{q_update_app} and the improved step size selection result in a substantial reduction of the number of iterations required over the number required using the choices made in \cite{TransmitterOptimization}.
Furthermore, we have observed that $\hat{\mathbf{Q}}^{1}$ and the corresponding matrix $\hat{\mathbf{Q}}^{n}$ at the termination of the algorithm are typically closely related. If that relationship can be determined with reasonable accuracy, this observation suggests that a predictive step could be used to further reduce the number of iterations. As an example of what can be done, in Section~\ref{sect_sim} we illustrate how replacing $\hat{\mathbf{Q}}^{1}$ with a simple affine prediction, $\hat{\mathbf{Q}}^{1}_{p}$, of the terminating matrix  $\hat{\mathbf{Q}}^{n}$ results in substantial reduction in the number of iterations.

%

\section{Conventional ZF beamforming with per-antenna power constraints} \label{sect_ZF}

Even though the computational cost of each iteration of the PAPCed offset maximization beamforming algorithms in the previous section is dominated by terms that are only $\mathcal{O}(N_t^3)$, when the BS has a large number of antennas the resulting computational load can still be substantial. The dominating components arise from determining the beamforming directions, and the fact that these directions are updated at each iteration. That suggests that we may be able to develop lower cost algorithms for systems with a large number of antennas if we could find a way to simplify the computation of the beamforming directions. In this section we will do that by developing variants of the nominal ZF directions, and we will integrate them with the robust power loading technique while ensuring that the required PAPCs are satisfied. In the following section we will develop analogous techniques based on variants of the MRT directions. For the ZF case, the beamforming directions are obtained using techniques developed in \cite{Zero-ForcingPrecoding}, but in the MRT case, the design of the beamforming directions appears to be new.

To develop PAPCed variants of the conventional ZF and MRT beamformers, we observe that in contrast to QoS-based designs, in which the SINR is controlled directly (e.g.,  \eqref{pr_gen_sinr}), the conventional ZF and MRT designs focus on the desired signal power and interference components of the SINR separately. In particular, given that the SINR for user $k$ is $\text{SINR}_k = \frac{\mathbf{h}_k^H \mathbf{w}_k \mathbf{w}_k^H \mathbf{h}_k}{\mathbf{h}_k^H  (\sum_{j \neq k}\mathbf{w}_j \mathbf{w}_j^H) \mathbf{h}_k + \sigma_k^2}$, if we were to maximize the minimum nominal received signal power subject to a total power constraint (i.e.,  $\max_{\substack{\{\mathbf{w}_k}\}}  \min_{\substack{k}} \mathbf{h}_{e_k}^H \mathbf{w}_k \mathbf{w}_k^H \mathbf{h}_{e_k}$ subject to $\sum_k \mathbf{w}_k^H \mathbf{w}_k \leq P_t$) we would obtain beamformers that are a particular power loading of the nominal MRT directions. If we were to add the nominal ZF  constraints on the interference into that problem  (i.e., $\mathbf{h}_{e_j}^H \mathbf{w}_k \mathbf{w}_k^H \mathbf{h}_{e_j}  = 0,  \forall k \neq j$), then we would obtain beamformers that are a particular power loading of the ZF directions  \cite{Zero-ForcingPrecoding}. Due to the structure of the total power constraint, in many simple beamforming problems the optimization of the beamforming directions decouples from the power loading. That is indeed the case for our formulation for MRT and ZF beamforming directions. As an example, if we were to maximize the minimum value of $\mathbf{h}_{e_k}^H \mathbf{w}_k \mathbf{w}_k^H \mathbf{h}_{e_k}/\|\mathbf{h}_{e_k}\|^2$, which is the power of the signal transmitted in the direction of user $k$, rather than the power received by that user, we would obtain a set of beamformers in the MRT or ZF directions, but with a different power loading.


When the total power constraint is replaced by PAPCs, the optimization of the beamforming directions becomes coupled with the power loading and hence the choice of the metric to optimize changes both the power loading and the directions. While our approach will work for either metric, and indeed for several others, we will focus on the second metric $\mathbf{h}_{e_k}^H \mathbf{w}_k \mathbf{w}_k^H \mathbf{h}_{e_k}/\|\mathbf{h}_{e_k}\|^2$. The rationale for this choice is that while the received signal power is suitable for the ZF problem in the perfect CSI case,  where the ZF constraints will eliminate the interference \cite{Zero-ForcingPrecoding}, it can be quite sensitive to the interference incurred due to channel estimation errors. (This is illustrated in our simulation results in Section~\ref{sect_sim}.) Accordingly, we define the normalized channel directions $\mathbf{h}_{n_k}=\mathbf{h}_{e_k}/\|\mathbf{h}_{e_k}\|$ and we formulate the following generic problem to obtain PAPCed versions of the conventional beamformers
\begin{subequations}\label{pa_MF_gen}
\begin{align}
     \max_{\substack{\mathbf{w}_k}, t} \quad & t \\
    \text{s.t.} \quad & \Bigl[\textstyle\sum_{k=1}^K \mathbf{w}_k \mathbf{w}_k^H \Bigr]_{i,i}\leq p_i, \quad \forall i \label{pa_MF_pa} \\
    &\mathbf{h}_{n_k}^H \mathbf{w}_k \mathbf{w}_k^H \mathbf{h}_{n_k}  \geq t \quad \forall k, \label{pa_MF_gen_sig} \\
    &\mathbf{h}_{n_j}^H \mathbf{w}_k \mathbf{w}_k^H \mathbf{h}_{n_j}  \leq \varepsilon \quad \forall k \neq j. \label{pa_MF_gen_int}
    \end{align}
\end{subequations}
The  value of $\varepsilon$ determines whether the problem is of the ZF type, the MRT type, or a variant thereof. When $\varepsilon$ is negligible compared to the noise power,  the formulation describes a ZF-based approach, and when $\varepsilon$ is of the order of the noise power this represents a regularized ZF-based approach; cf. \cite{Zeroforcingmethods}. When $\varepsilon$ is sufficiently large, the constraints in \eqref{pa_MF_gen_int} become inactive, and accordingly the formulation describes an  MRT-based approach.

One strategy for solving \eqref{pa_MF_gen} is to employ a semidefinite relaxation \cite{SemidefiniteRelaxation}. As in related beamforming methods based on semidefinite relaxation (e.g., \cite{Reference2}), that approach involves the solution of a convex optimization problem for a set of matrices and a post-processing step that extracts good beamformers from these matrices. However, the computational cost of solving the convex optimization problem is  even higher than that of the offset maximization algorithm, and that is only the cost of determining the beamforming directions. Accordingly, in the following sections we will present low-cost algorithms for robust beamforming with PAPCed variants of the ZF and MRT beamforming directions.

\subsection{ZF beamforming with PAPCs only}

When $\varepsilon=0$, the problem in \eqref{pa_MF_gen} involves finding the beamforming vectors that remove the interference at the receivers under the nominal channel conditions and satisfy the PAPCs. The essence of this problem was addressed in \cite{Zero-ForcingPrecoding} using a re-parametrization technique. In particular, let us define the matrix $\mathbf{H}$ as the matrix whose $k$th column is $\mathbf{h}_{n_k}$ and the matrix $\tilde{\mathbf{U}}_{\text{ZF}}= \mathbf{H} (\mathbf{H}^H \mathbf{H})^{-1}$. The $k$th column of $\tilde{\mathbf{U}}_{\text{ZF}}$, denoted $\tilde{\mathbf{u}}_{\text{ZF}_k}$ is a zero-forcing direction for the $k$th user with a unit signal gain; i.e., $\tilde{\mathbf{u}}_{\text{ZF}_k}^H \mathbf{h}_{n_k}=1$. If we let $\mathbf{H}_\perp $ denote a matrix whose columns form a basis for the null space of  $\mathbf{H}$, then the set of all ZF directions for the $k$th user is given by the $k$th column of $\tilde{\mathbf{U}}_{\text{ZF}} + \mathbf{H}_\perp  \mathbf{M}$, for an arbitrary scaling matrix $\mathbf{M}$. Accordingly, the solution to the problem in \eqref{pa_MF_gen} takes the form
\begin{equation}\label{zf_dir}
    \mathbf{w}_k = \sqrt{t} (\tilde{\mathbf{u}}_{\text{ZF}_k} + \mathbf{H}_\perp \mathbf{m}_k),
\end{equation}
where $\mathbf{m}_k$ is the $k$th column of matrix $\mathbf{M}$ \cite{Zero-ForcingPrecoding}. Note that the constraints in \eqref{pa_MF_gen_sig} and \eqref{pa_MF_gen_int} (with $\varepsilon=0$) are automatically satisfied by designing the precoding vectors $\mathbf{w}_k$ in the form in \eqref{zf_dir}. The conditions that remain to be met are the PAPCs, and that can be done by adjusting the scaling matrix $ \mathbf{M}$. In \cite{Zero-ForcingPrecoding}, this problem was formulated as a convex quadratically-constrained  program that can be efficiently solved
\begin{subequations}\label{Shlomo}
\begin{align}
     \min_{\substack{\mathbf{M}}, \hat{p}} \quad & \hat{p}  \\
    \text{s.t.} \quad & \| (\tilde{\mathbf{U}}_{\text{ZF}} +\mathbf{H}_\perp \mathbf{M})^H \tilde{\mathbf{e}}_i   \|^2  \leq \hat{p}, \quad \forall i, \label{Shlomo_papc}
     \end{align}
\end{subequations}
where $\tilde{\mathbf{e}}_i$ is the $i$th column of the identity matrix.  To complete the design, we choose the largest value for $t$ such that $t \hat{p} \leq p_i, \forall i$, which means that the beamformers of the form in \eqref{zf_dir} satisfy the remaining constraints; i.e., those in \eqref{pa_MF_pa}.

If we let the $(k,k)$th  entry of the diagonal matrix $\hat{\mathbf{Q}}$ denote the dual variable of the $k$th PAPC in \eqref{Shlomo_papc}, then the KKT conditions of the dual problem of \eqref{Shlomo} show that the scaling matrix should satisfy $\mathbf{M} = - ( \mathbf{H}_\perp^H \hat{\mathbf{Q}} \mathbf{H}_\perp)^{\dag}  (\mathbf{H}_\perp^H  \hat{\mathbf{Q}} \tilde{\mathbf{U}}_{\text{ZF}})$. Although such a relation does not allow for a closed-form solution, as we do not know $\hat{\mathbf{Q}}$, it does allow for the integration of the robust power loading method in \cite{Peruseroutageconstrained}, as an alternative to giving all the users the same nominal signal strength $t$. Furthermore, the explicit relation between $ \hat{\mathbf{Q}}$ and $\mathbf{M}$ allows us to use the sub-gradient algorithm for  $ \hat{\mathbf{Q}}$ and to calculate $\mathbf{M}$ accordingly. The proposed algorithm is summarized as Algorithm~\ref{Alg4}.

\begin{algorithm}
\caption{ZF with PAPCs and robust power loading}
\label{Alg4}
\begin{algorithmic}[1]
\State  Find $\mathbf{H}_\perp$ and $\tilde{\mathbf{U}}_{\text{ZF}}$. Initialize $ \hat{\mathbf{Q}}^0=\mathbf{I}$. Set $n=0$.
\While {$[\sum_{k} \mathbf{w}_k \mathbf{w}_k^H]_{i,i}-p_i>\epsilon_i$ for any $i$}
 \State Compute $\mathbf{M} = - ( \mathbf{H}_\perp^H \hat{\mathbf{Q}}^{n} \mathbf{H}_\perp)^{\dag}  (\mathbf{H}_\perp^H  \hat{\mathbf{Q}}^{n} \tilde{\mathbf{U}}_{\text{ZF}}).$
\State Find the beamformers directions $\{\mathbf{u}_k\}$  by normalizing $ \tilde{\mathbf{u}}_{\text{ZF}_k} + \mathbf{H}_\perp \mathbf{m}_k$.
  \State Find  $\{\beta_k\}$ and $r^\star$ by solving $\mathbb{E} (\mathbf{h}_{k}^H \mathbf{Q}_{k} \mathbf{h}_{k} - \sigma_{k}^2 )= \sigma_{s_k} r^\star$ and \eqref{extra_eqn} using the method provided in Section~\ref{power_loading_algo}.
  \State Update  $\hat{\mathbf{Q}}^{n+1}$  using the results in Appendix~\ref{q_update_app}.
  \State Increment $n$.
\EndWhile
\end{algorithmic}
\end{algorithm}

From a computational respective, the key steps in the initialization of this algorithm are the finding of the ZF directions and the null space of $\mathbf{H}$, which requires $\mathcal{O}(N_t^2 K)$ operations. Each iteration of the algorithm involves the iterative solution of the $K+1$ linear equations in step 5, which, as explained in Section~\ref{power_loading_algo}, requires $\mathcal{O}(K^3)$ operations, and the matrix operations required to update $\mathbf{M}$ in step 3, which require $\mathcal{O}((N_t-K)^3)$ operations.
When the number of antennas $N_t$ is close to the number of users $K$, the dimensions of the matrix  $\mathbf{H}_\perp^H \hat{\mathbf{Q}} \mathbf{H}_\perp$ are small, which means that in that case the computational cost of this algorithm is dominated by finding the ZF directions and the null space in the initialization step.

\subsection{Generalized ZF beamforming} \label{subsect_GZF}

The extension of the ZF design with PAPCs to accommodate a total power constraint is straightforward, and follows the same steps that were used in the generalized offset maximization problem; see Section~\ref{gen_offset_max}. The generalized ZF problem can be formulated by adding the total power constraint $\sum_{k=1}^K \mathbf{w}_k^H \mathbf{w}_k \leq P_t$ to the constraints in \eqref{pa_MF_gen}. Then we consider the equivalent power minimization problem, assuming, for now, that the optimal $t$ is known
\begin{subequations}\label{GZF_eq}
\begin{align}
     \min_{\substack{\mathbf{M}}} \quad & t \sum_i \| (\tilde{\mathbf{U}}_{\text{ZF}} +\mathbf{H}_\perp \mathbf{M}) \tilde{\mathbf{e}}_i   \|^2  \\
    \text{s.t.} \quad & t \| (\tilde{\mathbf{U}}_{\text{ZF}} +\mathbf{H}_\perp \mathbf{M})^H \tilde{\mathbf{e}}_i   \|^2 \leq p_i, \quad \forall i.
     \end{align}
\end{subequations}
Consistent with our previous analysis, we will let $\hat{\mathbf{Q}}$ denote the diagonal matrix with the dual variables of the PAPCs on its diagonal. From the KKT conditions  we can then show that $\mathbf{M} = - ( \mathbf{H}_\perp^H (\hat{\mathbf{Q}}+\mathbf{I}_{N_t}) \mathbf{H}_\perp)^{-1}  (\mathbf{H}_\perp^H  (\hat{\mathbf{Q}}+\mathbf{I}_{N_t}) \tilde{\mathbf{U}}_{\text{ZF}}).$  Furthermore, as in the previous algorithm we replace the uniform power loading, $t$, with the robust power loading from \cite{Peruseroutageconstrained}. The resulting modified version of Algorithm~\ref{Alg4} is stated in Algorithm~\ref{Alg5}. As is apparent from Algorithm~\ref{Alg5}, the order of its computational cost is the same as that of Algorithm~\ref{Alg4}.

\begin{algorithm}
\caption{Generalized ZF}
\label{Alg5}
\begin{algorithmic}[1]
\State  Find $\mathbf{H}_\perp$ and $\tilde{\mathbf{U}}_{\text{ZF}}$. Initialize $ \hat{\mathbf{Q}}^0=\mathbf{0}$. Set $n=0$.
\While {$[\sum_{k} \mathbf{w}_k \mathbf{w}_k^H]_{i,i}-p_i>\epsilon_i$ for any $i$}
\State Compute $\mathbf{M} = - ( \mathbf{H}_\perp^H (\hat{\mathbf{Q}}^{n}+\mathbf{I}_{N_t}) \mathbf{H}_\perp)^{-1}  (\mathbf{H}_\perp^H  (\hat{\mathbf{Q}}^{n}+\mathbf{I}_{N_t}) \tilde{\mathbf{U}}_{\text{ZF}}).$
\State Find the beamformers directions $\{\mathbf{u}_k\}$  by normalizing $ \tilde{\mathbf{u}}_{\text{ZF}_k} + \mathbf{H}_\perp \mathbf{m}_k$.
  \State Find $\{\beta_k\}$ and $r^\star$ by solving $\mathbb{E} (\mathbf{h}_{k}^H \mathbf{Q}_{k} \mathbf{h}_{k} - \sigma_{k}^2 )= \sigma_{s_k} r^\star$ and $\sum_k \beta_k = P_t$ using the method provided in Section~\ref{power_loading_algo}.
  \State Update $\hat{\mathbf{Q}}^{n+1}$ using \eqref{qupdate_gen}.
  \State Increment $n$.
\EndWhile
\end{algorithmic}
\end{algorithm}

\section{Conventional MRT with per-antenna power constraints} \label{sect_MRT}

As we have seen in the previous section, our approach to imposing PAPCs on the class of ZF beamformers can result in an algorithm of lower computational cost than that of offset maximization with PAPCs. However, any advantage is dependent on the size of the null space of the channel matrix. In settings with a large number of antennas and a small number of users, such as those arise in massive MIMO, the size of the null space can be quite large. In this section, we will show how the complexity can be further reduced by using an MRT-based approach rather than the ZF-based approach.

\subsection{MRT with PAPCs}

In the MRT case, the interference conditions $\mathbf{h}_{n_j}^H \mathbf{w}_k \mathbf{w}_k^H \mathbf{h}_{n_j}  \leq \varepsilon $ are omitted from the problem in~\eqref{pa_MF_gen}, and the problem of finding nominal MRT-based beamformers that satisfy PAPCs can be written as
\begin{subequations}\label{pa_MF_relaxed}
\begin{align}
     \max_{\substack{\mathbf{w}_k}, t} \quad & t \\
    \text{s.t.} \quad & \Bigl[\textstyle\sum_{k=1}^K  \mathbf{w}_k \mathbf{w}_k^H \Bigr]_{i,i}\leq p_i, \quad \forall i  \label{pa_MF_relaxed_papc} \\
    &\mathbf{h}_{n_k}^H  \mathbf{w}_k \mathbf{w}_k^H \mathbf{h}_{n_k}  \geq t, \quad \forall k. \label{pa_MF_relaxed_MF}
    \end{align}
\end{subequations}
Following a similar analysis to those performed earlier, if we let $q_i$ denote the dual variable for the $i$th PAPC, define the diagonal matrix $\hat{\mathbf{Q}}$ such that $[\hat{\mathbf{Q}}]_{i,i}=q_i$, and define $\nu_k$ to be the dual variable for the $k$th condition in \eqref{pa_MF_relaxed_MF}, then the Lagrangian of the problem in \eqref{pa_MF_relaxed} can be written as:
\begin{multline}
     \mathcal{L}(t,\mathbf{w}_k,\nu_k,q_i)= -t + \sum_{i=1}^{N_t}q_i \Bigl(\Bigl[\sum_{k=1}^K \mathbf{w}_k \mathbf{w}_k^H \Bigr]_{i,i}- p_i \Bigr) \\ -\sum_{k=1}^{K}\nu_k (\mathbf{h}_{n_k}^H \mathbf{w}_k \mathbf{w}_k^H  \mathbf{h}_{n_k} - t).
\end{multline}
Accordingly, we can state the KKT conditions in a simplified form as $\sum_{k=1}^{K}\nu_k =1$, $t = \sum_{i=1}^{N_t} q_i p_i$, and $\hat{\mathbf{Q}}\mathbf{w}_k = \nu_k \mathbf{h}_{n_k} \mathbf{h}_{n_k}^H \mathbf{w}_k.$ The last condition can be re-written as $\mathbf{w}_k = \nu_k  \mathbf{h}_{n_k}^H \mathbf{w}_k \hat{\mathbf{Q}}^{-1}\mathbf{h}_{n_k},$ which means that $\mathbf{w}_k $ and $\hat{\mathbf{Q}}^{-1}\mathbf{h}_{n_k}$ have the same direction. We note that at optimality $[\hat{\mathbf{Q}}]_{i,i}$ is equal to the $i$th element of $ \mathbf{h}_{n_k} $, scaled by $\nu_k \mathbf{h}_{n_k}^H \mathbf{w}_k,$ then divided by the $i$th element of $\mathbf{w}_k$. This equation does not allow any optimal $[\hat{\mathbf{Q}}]_{i,i}$ to be zero except if the channel vector $ \mathbf{h}_{n_k} $ contains a zero, which, under most reasonable channel models, is a ``zero-probability" event. Since each $q_i$ is positive, the constraints in \eqref{pa_MF_relaxed_papc} are all active, and accordingly $\textstyle\sum_k \beta_k= \textstyle\sum_i  p_i$.

Similar to the analysis of the previous problems, if we know $\hat{\mathbf{Q}}$, then we can find the beamforming directions using $\hat{\mathbf{Q}}^{-1}\mathbf{h}_{n_k}$ and subsequently solve for the power loading using the linear equations that arise when \eqref{pa_MF_relaxed_MF} holds with equality. The value of $t$ in \eqref{pa_MF_relaxed_MF} can be calculated using the KKT equation $t = \sum_{i=1}^{N_t} q_i p_i$.  If the equations in \eqref{pa_MF_relaxed_MF} were not satisfied with equality at optimality, we could rescale the beamforming vectors to get a larger value of $t$, which would contradict the assumed optimality. This observation is similar to the observation in the offset maximization section that enabled the use of the subgradient algorithm to find $\hat{\mathbf{Q}}$. Accordingly, we can suggest the iterative algorithm in Algorithm~\ref{Alg6}.

\begin{algorithm}
\caption{Nominal MRT with PAPCs}
\label{Alg6}
\begin{algorithmic}[1]
\State  Initialize $\hat{\mathbf{Q}}^{0} = \mathbf{I}$. Set $n=0$.
\While {$[\sum_{k} \mathbf{w}_k \mathbf{w}_k^H]_{i,i}-p_i>\epsilon_i$ for any $i$}
 \State Solve for the directions using $(\hat{\mathbf{Q}}^n)^{-1}\mathbf{h}_{n_k}$.
  \State Find the beamformer magnitudes $\{\beta_k \}$ and $t$ using the linear equations that arise when the constraints in \eqref{pa_MF_relaxed_MF} are satisfied with equality and $\textstyle\sum_k \beta_k= \textstyle\sum_i  p_i$.
  \State Update $\hat{\mathbf{Q}}^{n+1}$ using Appendix~\ref{q_update_app}.
  \State Increment $n$.
\EndWhile
\end{algorithmic}
\end{algorithm}

Algorithm~\ref{Alg6} provides an iterative way to find the values of $\hat{\mathbf{Q}}$, and, accordingly, the optimal precoding vectors. Its complexity per iteration is no more than linear in $N_t$. However, we will now develop a closed-form expression that approximates the optimal solution of Algorithm~\ref{Alg6} when the PAPCs are the same; i.e., $p_i=p, \forall i$. This closed-form removes the need for any iterations, which allows for an algorithm that is suitable for massive MIMO settings. To develop the approximation,  we first note that the PAPCs and the MRT constraints hold with equality at optimality. That means that at optimality
\begin{equation}\label{mrt_kkt}
\begin{aligned}
     \Bigl[\sum_{k=1}^K \mathbf{w}_k \mathbf{w}_k^H \Bigr]_{i,i} &  = \sum_{k=1}^K  \nu_k^2  | \mathbf{h}_{n_k}^H \mathbf{w}_k|^2  \left|\bigl[\hat{\mathbf{Q}}^{-1}\mathbf{h}_{n_k}\bigr]_{i}\right|^2 \\
     & = \sum_{k=1}^K  \nu_k^2 t  \left|\bigl[ \hat{\mathbf{Q}}^{-1}\mathbf{h}_{n_k}\bigr]_{i}\right|^2. \\
     & = p.
    \end{aligned}
\end{equation}
Now let us define
\begin{equation}\label{mrt_kkt2}
    g_i=\sum_{k=1}^K \nu_k^2  \left|\bigl[\mathbf{h}_{n_k}\bigr]_{i}\right|^2.
\end{equation}
Using \eqref{mrt_kkt}, we can also write  $g_i= p q_i^2/t = p q_i^2/(\sum_{i=1}^{N_t} q_i p).$
Accordingly, we can calculate $q_i$ from $\{g_i\}$ as  $q_i= (\sum_j \sqrt{g_j}) \sqrt{g_i}.$
The objective of maximizing $t = \sum_{i=1}^{N_t} q_i p$ is, therefore, equivalent to maximizing $\sum_j \sqrt{g_j}$.  Since the dual variables $\nu_k^2$ enter \eqref{mrt_kkt2} as weighting variables for the power gains of the components of $\mathbf{h}_{n_k}$, the optimal  values of $\nu_k$ are influenced by the relative values of the elements of each set $\{ \left|\bigl[\mathbf{h}_{n_k}\bigr]_{i}\right|^2 \}_{k=1}^K$. When these elements have the same distribution, the optimal values of  $\nu_k$ tend to get closer as the number of antennas grows. Since $\sum_k  \nu_k=1$, that suggests the approximation $\nu_k\approx 1/K$. Since the approximation only holds in the limit, there will be discrepancy between the actual power on the antennas and $p$, but as the number of antennas grows, that difference decreases. For a finite number of antennas, we may rescale the result so that  the PAPCs are satisfied. That is done in steps 6 and 7 in  Algorithm~\ref{Alg7}.

\begin{algorithm}
\caption{One-shot approximate nominal MRT with PAPCs}
\label{Alg7}
\begin{algorithmic}[1]
 \State Approximate $\nu_k \approx 1/K$.
  \State Calculate $g_i=\sum_{k=1}^K \nu_k^2  \left|\bigl[\mathbf{h}_{n_k}\bigr]_{i}\right|^2$ and $t=\sum_j \sqrt{g_j}$.
  \State Calculate $\hat{\mathbf{Q}}$ using $q_i= (\sum_j \sqrt{g_j})  \sqrt{g_i}$.
  \State Calculate the beamformer directions $\{\mathbf{u}_k\}$ by normalizing  $\hat{\mathbf{Q}}^{-1}\mathbf{h}_{n_k}$.
  \State Find $\beta_k=t/\mathbf{h}_{n_k}^H \mathbf{u}_k \mathbf{u}_k^H \mathbf{h}_{n_k} $.
  \State Form the vector $\mathbf{y}$, such that $y_i=[\sum_{k=1}^K \mathbf{w}_k \mathbf{w}_k^H \Bigr]_{i,i}$.
  \State Form the correction vector $\mathbf{z}$ such that $z_i=\sqrt{p_i/y_i}$.
  \State Correct each beamformer vector by element-wise multiplying each  $\mathbf{w}_k$ by the correction vector $\mathbf{z}$.
\end{algorithmic}
\end{algorithm}

Both of the algorithms for the nominal MRT-based approach (Algos \ref{Alg6} and \ref{Alg7})  result in beamformers that satisfy the PAPCs. As we will see in the simulation section, the resulting beamformers provide similar outage performance even for relatively small number of antennas. However, both of the algorithms are based on nominal performance criteria and any robustness that is obtained arises only implicitly. To address that point, we observe that Algorithm~\ref{Alg6} updates $\hat{\mathbf{Q}}$ iteratively using the sub-gradient algorithm, which allows for the incorporation of the robust power loading described in Section~\ref{power_loading_algo}. The ability to incorporate that power loading can significantly reduce the outage probability by allocating each user an appropriate amount of power rather than forcing the nominal signal power of different users to be the same value $t$. The resulting algorithm is stated in Algorithm~\ref{Alg8}. In scenarios in which it is reasonable to use the same value of $t$ for all users, or when we can pre-define different weights for the value of $t$, Algorithm~\ref{Alg7} can provide a closed-form solution that is close to the optimal one, without the need for any iterations.

\begin{algorithm}
\caption{Robust MRT with PAPCs}
\label{Alg8}
\begin{algorithmic}[1]
\State Initialize $\hat{\mathbf{Q}}^{0} = \mathbf{I}$. Set $n=0$.
\While {$[\sum_{k} \mathbf{w}_k \mathbf{w}_k^H]_{i,i}-p_i>\epsilon_i$ for any $i$}
  \State Find the beamforming directions $\{\mathbf{u}_k\}$ using  $\hat{\mathbf{Q}}^{-1}\mathbf{h}_{n_k}$.
  \State  Find $\{\beta_k\}$ and $r^\star$  by solving $\mathbb{E} (\mathbf{h}_{k}^H \mathbf{Q}_{k} \mathbf{h}_{k} - \sigma_{k}^2 )= \sigma_{s_k} r^\star$ and \eqref{extra_eqn} using the method provided in Section~\ref{power_loading_algo}.
  \State Update $\hat{\mathbf{Q}}^{n+1}$ using Appendix~\ref{q_update_app}.
  \State Increment $n$.
\EndWhile
\end{algorithmic}
\end{algorithm}

The complexity of Algorithm~\ref{Alg8} is dominated by operations that are linear in the number of antennas for each user. This means that the complexity per iteration is of the order of $\mathcal{O}(N_t K)$ operations. The robust power loading can be effectively approximated in the massive MIMO settings so that it requires only $\mathcal{O}(N_t K)$ operations, beside the $\mathcal{O}(K^3)$ operations for the initial matrix inversion \cite{Peruseroutageconstrained}.

\subsection{Generalized MRT}

The derivation of the MRT-based algorithm when the total power constraint is added to \eqref{pa_MF_relaxed} follows the same steps that were performed in the ZF case and the offset maximization case. The modified algorithm is presented in Algorithm~\ref{Alg9}.

\begin{algorithm}
\caption{Generalized MRT}
\label{Alg9}
\begin{algorithmic}[1]
\State  Set $\hat{\mathbf{Q}}^{0}=\mathbf{0}$, and $n=0$.
\While {$[\sum_{k} \mathbf{w}_k \mathbf{w}_k^H]_{i,i}-p_i>\epsilon_i$ for any $i$}
  \State Find the beamformers directions $\{\mathbf{u}_k\}$ by normalizing  $(\mathbf{I}_{N_t}+\hat{\mathbf{Q}})^{-1}\mathbf{h}_{n_k}$.
  \State  Find $\{\beta_k\}$ and $r^\star$  by solving $\mathbb{E} (\mathbf{h}_{k}^H \mathbf{Q}_{k} \mathbf{h}_{k} - \sigma_{k}^2 )= \sigma_{s_k} r^\star$ and $\sum_k \beta_k = P_t$ using the method provided in Section~\ref{power_loading_algo}.
  \State Update $\hat{\mathbf{Q}}^{n+1}$ using \eqref{qupdate_gen}.
  \State Increment $n$.
\EndWhile
\end{algorithmic}
\end{algorithm}

\section{Simulation results}\label{sect_sim}

In this section, we will show how the application of PAPCs to substantially reduce the dynamic range of the power transmitted from each antenna can be implemented without significantly degrading the outage probability of the system. We consider a system in which a BS with $N_t$ antennas serves $K$ single-antenna users distributed uniformly in a disk of radius 3.2km around  the BS. The large scale fading is modelled using a path-loss exponent of 3.52 and log-normal shadow fading with 8dB standard deviation. The small scale fading is modelled using the standard i.i.d. Rayleigh model. We assume an additive channel estimation error of covariance $0.04 \mathbf{I}$, and an SINR target of $\gamma=3$dB for all users. For the algorithms with PAPCs only, the PAPC is uniform and is set to $p_i=P_t/N_t$, where $P_t$ is the total power constraint, which is implicit in this case. For the generalized algorithms with both PAPCs and a total power constraint (Algos 3, 5, and 9), the PAPCs are set to be slightly larger, so that the total power constraint is active. For these cases we choose $p_i=1.2 P_t/N_t$. We assume that each user has a signal sensitivity of -90dBm, and we will consider this power as the noise power. The termination parameter for the algorithms is chosen to be $\epsilon_i=0.1p_i$, and each experiment is repeated on 20,000 channel realizations. A simple channel-strength user selection technique is employed, where users having $\|\mathbf{h}_{e_k}\|^2 P_t/ k\sigma_k^2 \geq \gamma_k$ are served.

\begin{figure}
\begin{center}
    \epsfysize= 2.8in
     \epsffile{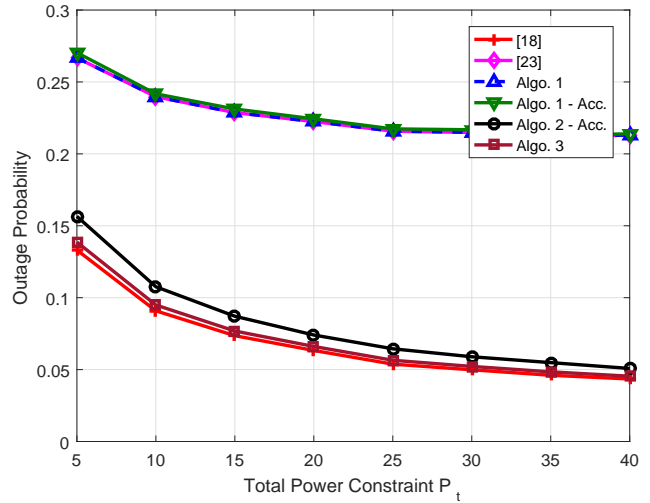}
\caption{Outage probability for a 4 antenna BS serving 3 users with a total transmitted power of $P_t$.
}\label{pa_fig1}
\end{center}
\end{figure}

To demonstrate the application of PAPCs with offset maximization, in Fig.~\ref{pa_fig1} we plot the outage probability versus the total power constraint $P_t$ for six different algorithms in a scenario in which $N_t=4$ and $K=3$. The first algorithm is the nominal PAPCed design algorithm presented in~\cite{TransmitterOptimization}, with the beamforming vectors scaled so that the total power is equal to $P_t$. This is equivalent to solving~\eqref{pr_gen3} when  $r_{pa}^\star=0$, then scaling the resulting beamforming vectors. We compare the performance of~\cite{TransmitterOptimization} to the performance of Algo.~\ref{Alg1}  with and without the acceleration step, and Algo.~\ref{Alg2}  with the acceleration step. We note that while the performance of Algo.~\ref{Alg1} is close to that of~\cite{TransmitterOptimization}, the application of the robust power loading in Algo.~\ref{Alg2} provides a significant reduction in the outage probability.
To asses the impact of the PAPCs we compare the performance of  Algo.~\ref{Alg2} to that of the robust offset maximization technique with a total power constraint only~\cite{Peruseroutageconstrained}. As seen in Fig.~\ref{pa_fig1}, Algo.~\ref{Alg2} achieves a performance close to that of~\cite{Peruseroutageconstrained} even though it imposes PAPCs. As expected, the performance of Algo.~\ref{Alg3}, which imposes a total power constraint and weaker PAPCs,  falls in between that of~\cite{Peruseroutageconstrained} and  Algo.~\ref{Alg2}.

\begin{figure}
\begin{center}
    \epsfysize= 2.8in
     \epsffile{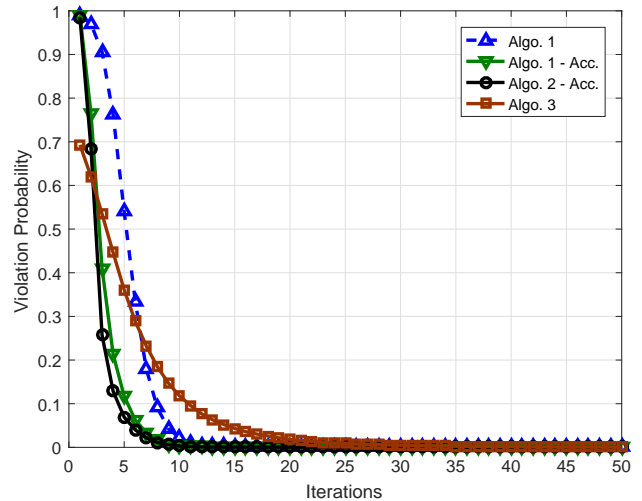}
\caption{Convergence behaviour for a 4 antenna BS serving 3 users with a total transmitted power of $P_t$. The violation probability measures the fraction of the  20,000  realizations for which at least one PAPC was violated by more then $10\%$ at the given iteration of the algorithm.
}\label{pa_fig2}
\end{center}
\end{figure}

The convergence rate of the subgradient algorithm strongly depends on how the step size is chosen and, hence, this should be tailored to the application. Based on insights from~\cite{DecompositionbyPartialLinearization}  we have chosen a step size that is updated using $t_n=t_{n-1}-t_{n-1}^2/1000$. Our numerical experience has suggested choosing $t_0=N_t/( P_t K)$. To examine the potential impact of the prediction scheme outlined in Section~\ref{algo_acc}, we have implemented a linear predictor of the form $\hat{\mathbf{Q}}^{1}_{p} = 2.8~\text{diag}(\mathbf{q}^1) - 1.8\mathbf{I}$. To show the effectiveness of these choices, we plot in Fig.~\ref{pa_fig2} the percentage of violated PAPCs versus the iteration number for the scenario in which $P_t=40$. We set the violation to one when any antenna is transmitting a power that is more than $10\%$ higher than $p_i$. (Recall that we set $\epsilon_i=0.1 p_i$.) We observe from Fig.~\ref{pa_fig2} that within the first few iterations, the PAPCs are met in most cases. We also note that the acceleration step can reduce the average number of iterations while providing almost the same outage performance. In order to provide context for these results, we point out that the average number of iterations required by the nominal algorithm in~\cite{TransmitterOptimization} is much higher. Indeed, as shown in~\cite{TransmitterOptimization}, it can range from a few tens to hundreds in analogous settings.

In assesing the performance of the ZF-based PAPCed beamforming algorithms, rather than examining the outage performance against the transmission power, we will fix the total power constraint to $P_t=2$ and examine the performance as the number of antennas, $N_t$, increases. Other than that, the scenario is the same as the previous one. As performance benchmarks for Algo.~\ref{Alg4}, we have included the performance of the algorithm in~\cite{Zero-ForcingPrecoding} which maximizes the minimum received signal power, $\mathbf{h}_{e_k}^H \mathbf{w}_k \mathbf{w}_k^H \mathbf{h}_{e_k}$, and a modified version of the algorithm in~\cite{Zero-ForcingPrecoding} that maximizes the minimum value of $\mathbf{h}_{n_k}^H \mathbf{w}_k \mathbf{w}_k^H \mathbf{h}_{n_k}$ instead.
We observe that in the case of noisy channel estimates, the normalization step significantly reduces the outage probability. More importantly, the application of the robust power loading in Algo.~\ref{Alg4} provides significantly better performance. As a lower bound on the outage achieved by Algo.~\ref{Alg4} we consider ZF beamforming with the nominal ZF directions and robust power loading with only a total power constraint~\cite{Peruseroutageconstrained}; i.e., without the PAPCs. The resulting comparison shows that the degradation incurred by imposing the PAPCs is quite small. Finally, as expected, the performance of the generalized algorithm (Algo.~\ref{Alg5}) lies in between that of Algo.~\ref{Alg4} and that of~\cite{Peruseroutageconstrained}.

\begin{figure}
\begin{center}
    \epsfysize= 2.8in
     \epsffile{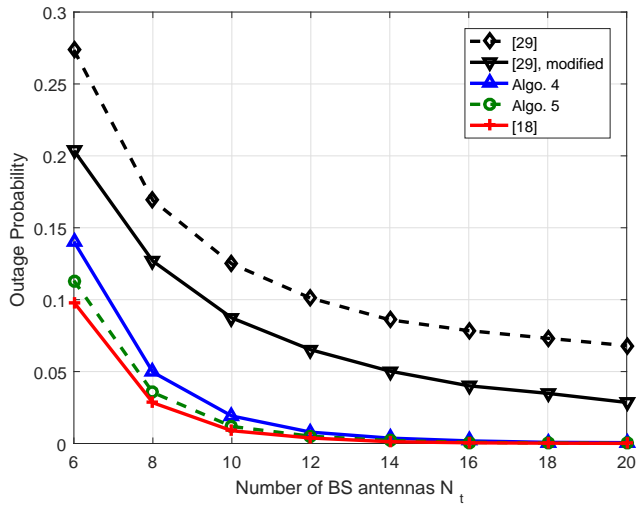}
\caption{Outage probability for a BS serving 3 users with a total transmitted power of $P_t=2$.
}\label{pa_fig3}
\end{center}
\end{figure}

%
%

\begin{figure}
\begin{center}
    \epsfysize= 2.8in
     \epsffile{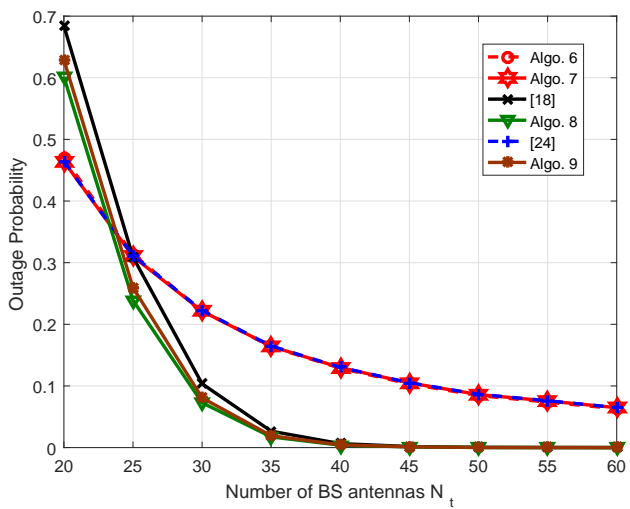}
\caption{Outage probability for a BS serving 8 users with a total transmitted power of $P_t=1$.
}\label{pa_fig5}
\end{center}
\end{figure}

To assess the performance of the MRT-based PAPCed algorithms, we will allow for more users, $K=8$, and set the total power constraint $P_t$ to be 1. As in the ZF case, we examine the outage performance versus the total number of antennas, $N_t$, but we do so for a larger number of antennas. In Fig.~\ref{pa_fig5}, the performance of Algos~\ref{Alg6},  \ref{Alg7}, and \ref{Alg8}  is compared to the performance of the algorithm in~\cite{ModifiedMRT}. We observe that the performance of  Algos~\ref{Alg6}, and \ref{Alg7} is almost identical to that of the algorithm presented in~\cite{ModifiedMRT}, and that the performance of Algo.~\ref{Alg8} is superior. As a benchmark, the performance of the robust MRT beamformer with only a total power constraint (i.e., no PAPCs)~\cite{Peruseroutageconstrained} is plotted in Fig.~\ref{pa_fig5}.  The performance of the generalized algorithm (Algo.~\ref{Alg9}) is also plotted.

%

\section{Conclusion}

In this paper, we developed low-complexity algorithms for finding robust beamformers that provide low outage of target SINRs while satisfying specified per-antenna power constraints (PAPCs). Initially, we used insights from the subgradient method for designing PAPCed beamformers in the case of perfect channel state information~\cite{TransmitterOptimization} to obtain PAPCed version of the offset maximization algorithm developed in~\cite{LowComplexityRobustMISO}. Further reductions in the outage probability were then obtained by incorporating the robust power loading presented in~\cite{Peruseroutageconstrained} into the design problem. While the proposed algorithms are of low complexity, we identified the evaluation of the beamforming directions as the computational bottleneck. To address that, we developed algorithms that employ PAPCed variants of the conventional zero-forcing (ZF) and maximum ratio transmission (MRT) directions and incorporate the robust power loading. In the process of doing so, we developed a closed-form expression for an MRT-based beamformer that satisfies PAPCs and may be appropriate for massive MIMO systems. Our simulation results revealed that PAPCed beamforming can be achieved without incurring a significant degradation in outage performance.

%

\appendices

\section{$\hat{\mathbf{Q}}$ update} \label{q_update_app}

To determine the updated value for $\hat{\mathbf{Q}}^{n+1}$, we have to determine the projection,
$$\hat{\mathbf{Q}}^{n+1}=\text{proj}\bigl(\hat{\mathbf{Q}}^{n}+ t_n \text{diag}(\text{diag} (\textstyle\sum_{i} \mathbf{w}_i \mathbf{w}_i^H))\bigr).$$
To do so, we let $\mathbf{q} = \text{diag}(\hat{\mathbf{Q}}^{n+1})$, and $\mathbf{q}_o = \text{diag}(\hat{\mathbf{Q}}^{n}+ t_n \text{diag} (\text{diag} (\textstyle\sum_{i} \mathbf{w}_i \mathbf{w}_i^H)))$. That enables us to write the projection problem as
\begin{subequations}\label{q_opt}
\begin{align}
      \min_{\substack{\mathbf{q}}} \quad & \| \mathbf{q} - \mathbf{q}_o \|^2 \\
   \text{s.t.} \quad & \textstyle\sum_{i=1}^{N_t} q_i p_i = \sum_{i=1}^{N_t} p_i \\
    & q_i \geq 0, \quad \forall i.
    \end{align}
\end{subequations}
If we let $\zeta$ denote the dual variable of the equality constraint, then from the KKT conditions of \eqref{q_opt} we can show that the optimal $q_i$ is
$$q_i = \max (q_{o_i} -   p_i \zeta/2,0),$$
where $\zeta/2=\bigl(\sum_{i, \forall q_i\neq 0}  p_i {q}_{o_i}- \sum_i  p_i \bigr) / \sum_{i, \forall q_i\neq 0}  p_i^2$. Given the nature of dependence of  $\{q_i\}$ and $\zeta$ on each other,  we will solve for their values using a fixed-point approach. First, we initialize $\zeta=0$, and then we iteratively calculate $q_i$ and $\zeta$ from the provided equations until their values stabilize.

In the case of  equal $p_i$  (i.e., $p_i=p, \forall i$), and when all the $q_i$ are positive  (i.e., all the PAPCS are active),  the update equation can be simplified to
\begin{equation}\label{qupdate}
 \hat{\mathbf{Q}}^{n+1}=\hat{\mathbf{Q}}^{n}+ t_n \text{diag} \bigl( \text{diag}  \bigl(\sum_{i} \mathbf{w}_i \mathbf{w}_i^H- p \mathbf{I}\bigr)\bigr).
\end{equation}

\end{document}